\documentclass[preprint,showpacs,preprintnumbers,amsmath,amssymb]{revtex4}
\usepackage{graphicx}
\usepackage{dcolumn}

\newcommand{\ff}{\mbox{\boldmath$f$}}

\newcommand{\bDelta}{\mbox{\boldmath$\Delta$}}

\newcommand{\xiprime} {\ensuremath{\boldsymbol{\xi}^{\prime}}}
\newcommand{\mxi}{\ensuremath{\boldsymbol{\xi}}}
\begin{document}
\title
{Kinetic boundary conditions in the lattice Boltzmann method}
\author{Santosh Ansumali}
 \affiliation{
ETH-Z\"urich, Department of Materials, Institute of Polymers\\
ETH-Zentrum, Sonneggstr. 3, ML J 19, CH-8092 Z\"urich, Switzerland
\\
}
\author{ Iliya V.\ Karlin}%
\affiliation{%
ETH-Z\"urich, Department of Materials, Institute of Polymers\\
ETH-Zentrum, Sonneggstr. 3, ML J 19, CH-8092 Z\"urich, Switzerland
\\
}%

\begin{abstract}
Derivation of the lattice Boltzmann method from
the continuous kinetic theory  [X.\ He and L.\ S.\ Luo,
{\it Phys.\ Rev.\ E} {\bf 55}, R6333 (1997); X.\ Shan and X.\ He,
{\it Phys.\ Rev.\ Lett.} {\bf 80}, 65 (1998)] is extended in order to obtain boundary conditions
for the method.  For the model of a diffusively reflecting moving
solid wall, the boundary condition for the discrete set of velocities
is derived, and the error of the discretization is
estimated. Numerical results are presented which demonstrate
convergence to  the hydrodynamic limit. In particular, the Knudsen
layer in the Kramers' problem is reproduced correctly for small
Knudsen numbers.
\end{abstract}
\maketitle

\section{Introduction}
In recent years, the lattice Boltzmann method (LBM) has emerged as an
alternative tool for the computational fluid dynamics
\cite{review}. 
Originally, the LBM was  developed as a modification of
the lattice gas model \cite{Frisch}. 
Later derivations \cite{ShanHe, Luo1}   revealed that the
method is  a special discretiszation  of the
 continuous Boltzmann equation. The derivation
  of the LBM \cite{ShanHe} from the Boltzmann equation is
 essentially  based on  Grad's moment method \cite{Grad}, together with the
  Gauss-Hermite quadrature in the velocity space.

 Another important
 issue was  to retain  positivity of discrete velocities populations
 in the bulk. Recently, a progress has been achieved in incorporating the
$H$--theorem into the method \cite{ K99, AK00, AK01},  and thus retaining
  positivity of the populations in the bulk. On the contrary, despite of several attempts \cite{Zieglar,Ladd94,  LuoBC, Noble,FD,
  NobleMod1, NobleMod2, NewBC} a fully consistent theory of the boundary
condition for the method is still lacking. It appears that the concerns
 about  positivity of the population, and the connection with the continuous
 case, are somewhat ignored while introducing the boundary condition. The
 way the no--slip condition for the moving wall is incorporated in the
 method \cite{Ladd94, Noble, NobleMod1}  is
 especially prone to danger of loss of positivity of the populations at
 the boundary. A clear understanding of
the boundary condition becomes  demanding  for the case of moving boundary,
complicated geometries,  chemically reactive or porous walls.

The theory of boundary conditions for  the
continuous Boltzmann equation is sufficiently well developed to incorporate
  the information about the structure and the chemical processes on the wall \cite{CBoltz}.
The
realization that the LBM is a special
discretization of the Boltzmann equation allows  to derive the
  boundary conditions for the
LBM from continuous kinetic theory.
In this work we demonstrate how this can be done in a systematic way.  
  
 The outline of the paper is as follows: In section \ref{Sec:1} we
 give a brief description  of the LBM. In section \ref{Sec:2} we
 briefly describe how boundary condition is formulated for the
 continuous kinetic theory. In section \ref{Sec:3} we
  derive the boundary condition for the  LBM and 
in section \ref{Sec:4} we
demonstrate some numerical simulation to validate the result.

\section{Overview of the method }
\label{Sec:1}     
In the LBM setup, one considers populations $f_i$ of discrete
velocities $\mathbf{c}_i$, where $i=1,\dots,b$, at discrete time
$t$. 
It is convenient to introduce $b$-dimensional
population vectors $\ff$.

In the isothermal case considered below,  local hydrodynamic variables are given as,
\begin{align}
\label{Hyd}
\begin{split} \rho= \sum_{i = 1}^{b} f_i(\mathbf{r}, t), \\
\rho \mathbf{u} 
= \sum_{i=1}^{b} \mathbf{c}_i f_i(\mathbf{r}, t).
\end{split}
\end{align}

The basic equation to be solved is
\begin{equation}
\label{LBM}
f_i(\mathbf{r}+\mathbf{c}_i,t+1)-f_i(\mathbf{r},t)= -\beta \alpha[\ff(\mathbf{r},t)]  
\bDelta_i[\ff(\mathbf{r},t)],
\end{equation}

 where  $\beta$ is a fixed parameter  in the
interval $[0,1]$ and is related to the viscosity.  A  scalar function  of the population
vector $\alpha$   is
the nontrivial root of the  nonlinear equation 
\begin{equation}
\label{step}
H(\ff)=H(\ff +\alpha\bDelta[\ff]).
\end{equation}
The function $\alpha$ ensures the discrete-time $H$--theorem.
In the previous derivations \cite{ShanHe,Luo1} of the LBM  from the Boltzmann equation, a quadratic
form for the equilibrium  distribution function $f^{\rm eq}$,
was  obtained by evaluating the Taylor series expansion of the
absolute Maxwellian equilibrium on the nodes of a properly selected
quadrature. This was done to ensure that 
 the Navier-Stokes equation is reproduced up to the order
$O(M^2)$, where $M$ is the Mach number.
However, the disadvantage of expanding equilibrium distribution
 function is that the condition of monotonicity of the  entropy production is not
 guaranteed. In order to avoid  this problem, in the entropic
 formulation \cite{K99, AK00, AK01}, the
 Boltzmann  $H$ function, rather than the equilibrium distribution, is
 evaluated at the nodes of the given quadrature, to get the
 discrete version of the $H$--function as
\begin{equation}
\label{result9}
H=\sum_{i=1}^{b}f_{i}\ln\left(
\frac{f_{i}}{w_i} \right),
\end{equation}
where  $w_i$ denotes the  weight associated with the
corresponding quadrature node $\mathbf{c}_i$. In the Appendix
\ref{app:1}, the derivation of the $H$--function is presented. 
Afterwards, the collision term
 is  constructed from the knowledge of the $H$--function (Eq.(\ref{result9})). 
 The  collision term $\bDelta$ is constructed in such a way
 that it satisfies 
a set of admissibility conditions  needed to have a proper $H$--theorem and
conservation laws (see Ref. \cite{AK01} for details).

  The   LBM  model with the BGK collision form \cite{Quin, ShanHe},
 can be considered as a limiting  case of the entropic
 formulation. To obtain the lattice BGK equation,  the function
 $\alpha$ in the Eq. (\ref{LBM}) is set 
 equals to $2$, and  for the  collision term $\bDelta$ BGK form is
 chosen. The equilibrium function used in the BGK form is obtained
 as the minimizer of the $H$--function (Eq.(\ref{result9}))  subjected
 to the hydrodynamic constrains (Eq.(\ref{Hyd})), evaluated
 up to the order $M^2$ \cite{K99}. 
Derivation of the boundary conditions  done in the subsequent section
 applies to both the forms of the LBM.

\section{ Boundary condition for the  Boltzmann equation}
\label{Sec:2}    
 Following  Ref. \cite{CBoltz},  we  briefly outline how boundary
 condition is formulated in the
 continuous kinetic theory. We shall restrict our discussion to the
case where the mass flux through the wall is zero. For the present
 purpose, a wall $\partial R$ is completely specified at any point $( \mathbf{r} \in\partial R )$ by the knowledge
 of the inward unit normal
 $\mathbf{n}$, the  wall
 temperature $T_{\mbox{w}}$ and the wall velocity $\mathbf{U}_{\mbox{w}}$. Hereafter, we  shall denote the distribution function in a frame of reference moving
 with the wall velocity as $g(\mxi)$, with
$ {\mxi}=  \mathbf{c} -\mathbf{U}_{\mbox{w}}$. 
The distribution function reflected from  the  non--adsorbing  wall can be written explicitly, if
the scattering probability is known.  
In explicit form,
\begin{align}
\label{probC}
 |\mxi \cdot\mathbf{n}|
g(\mxi, t)
 &= \int_{\xiprime \cdot \mathbf{n} < 0 } |\xiprime \cdot\mathbf{n}|
g(\xiprime, t)
 B\left(\mxi^\prime\rightarrow \mxi \right) d \xiprime
,
\; (\mxi \cdot \mathbf{n} > 0), 
\end{align}
where the  non--negative function  $ B \left( \xiprime \rightarrow \mxi
  \right )$ denotes
the  scattering probability from the direction $\mxi^\prime $ to the
direction $\mxi$. If the wall is  non-porous and 
 non-adsorbing, the total probability for an impinging particle to be
 re--emitted is unity:
\begin{align}
\label{norm}
\int_{\mxi \cdot \mathbf{n} > 0 } 
 B\left(\mxi^\prime\rightarrow
  \mxi \right) d \mxi = 1.
\end{align}
Eq. (\ref{probC}) and Eq. (\ref{norm}) ensure that the reflected distribution functions are
 positive and the normal flux through the wall is zero. 
A further restriction on the form of function
$B$ is dictated by the condition of detailed balance \cite{CBoltz},
\begin{equation}
\label{BPW}
 |\mxi^\prime \cdot\mathbf{n}|
g^{\mbox{eq}}(\mxi^\prime,\rho_{\mbox{w}}, 0, T_{\mbox{w}})
B\left(\mxi^\prime\rightarrow \mxi \right)
 = |\mxi \cdot\mathbf{n}|
g^{\mbox{eq}}(-\mxi,\rho_{\mbox{w}}, 0, T_{\mbox{w}})
B\left(-\mxi\rightarrow -\mxi^\prime \right).  
\end{equation}
A consequence of this property is that, if the impinging distributions
are wall--Maxwellian, then the reflected distributions are also wall--Maxwellian. Thus,
\begin{align}
 |\mxi \cdot\mathbf{n}|
g^{\mbox{eq}}(-\mxi, \rho_{\mbox{w}}, 0, T_{\mbox{w}})
 &= \int_{\mxi^\prime \cdot \mathbf{n} < 0 } |\mxi^\prime \cdot\mathbf{n}|
g^{\mbox{eq}}(\mxi^\prime, \rho_{\mbox{w}}, 0, T_{\mbox{w}})
 B\left(\mxi^\prime\rightarrow \mxi \right) d \mxi^\prime.
\end{align}
 This equation can also be understood as a weaker statement of the
 detailed balance condition \cite{CBoltz}. This form of the detailed
 balance is very attractive for our present purpose because of its
 integral nature, so that  a discretization can be done in a natural way.

 In this paper, we only
 consider   the diffusive boundary conditions because
 the steps associated with the discretization are 
 easier to appreciate due to the mathematical simplicity in this case.
In this model of the  wall it is assumed that the
out--going stream has completely lost its memory about the incoming
stream.  Thus, the scattering probability $B$ is independent of the
 impinging directions, and is equals to 
\begin{align}
B\left(\mxi^\prime\rightarrow \mxi \right) =
\frac{ |\mxi \cdot\mathbf{n}|
g^{\mbox{eq}}(-\mxi, \rho_{\mbox{w}}, 0, T_{\mbox{w}})
 }{ \int_{\mxi^\prime \cdot \mathbf{n} < 0 } |\mxi^\prime \cdot\mathbf{n}|
g^{\mbox{eq}}(\mxi^\prime, \rho_{\mbox{w}}, 0, T_{\mbox{w}}) d
\mxi^\prime}
\equiv B\left(\mxi\right)
.
\end{align}
Thus, the explicit expression for the reflected distribution function is
\begin{align}
g(\mxi,t)
 &=  \frac{
\int_{\mxi^\prime \cdot \mathbf{n} < 0 } |\mxi^\prime \cdot\mathbf{n}|
g(\mxi^\prime, t)
  d\mxi^\prime
}{
\int_{\mxi^\prime \cdot \mathbf{n} < 0 } |\mxi^\prime \cdot\mathbf{n}|
g^{\mbox{eq}}(\mxi^\prime, \rho_{\mbox{w}}, 0, T_{\mbox{w}} ) d\mxi^\prime} 
 g^{\mbox{eq}}(-\mxi, \rho_{\mbox{w}}, 0, T_{\mbox{w}} ),
\qquad (\mxi \cdot \mathbf{n}  > 0).
\end{align}

 We need to transform this equation into the stationary co--ordinate
system.  
As the equilibrium distribution depends only on
the difference between the particle velocity and the local velocity,
we have

\begin{align}
\label{CBC}
f(\mathbf{c},t)
 &=  \frac{
\int_{\mxi^\prime \cdot \mathbf{n} < 0 } |\mxi^\prime \cdot\mathbf{n}|
f(\mathbf{c}^\prime, t)
  d\mathbf{c}^\prime
}{
\int_{\mxi^\prime \cdot \mathbf{n} < 0 } |\mxi^\prime \cdot\mathbf{n}|
f^{\mbox{eq}}(\mathbf{c}^\prime, \rho_{\mbox{w}}, U_{\mbox{w}}, T_{\mbox{w}}) d \mxi^\prime}  
f^{\mbox{eq}}(\mathbf{c},\rho_{\mbox{w}}, U_{\mbox{w}}, T_{\mbox{w}}),
\qquad ( \left( \mathbf{c} -\mathbf{U}_{\mbox{w}}\right)\cdot  \mathbf{n}  > 0).
\end{align}

In the next section, we will show how the discretization of the
equation (\ref{CBC}) can be performed.  
\section{Discretization of the boundary condition }
\label{Sec:3}

 In the derivation of the lattice Boltzmann equation for the bulk various
integrals, which are evaluated at the nodes of a Gauss--Hermite quadrature,
\cite{Luo1} are of the form
\begin{equation}
\label{GHQ}
I = \int_{\mxi\in R^D} \exp{(- \mxi^2)} \phi(\mxi) d\mxi ,
\end{equation}
where $D$ is the spatial dimension.
This form of the integral is 
 well approximated by  the Gauss--Hermite quadrature.  However, the situation is 
different  on the boundary because  integrals appearing in 
 Eq. (\ref{CBC}) are over half--space.  The choice
of the  quadrature in the bulk was based on the properties of integrals in
the $R^D$. If we would  evaluate the integrals in Eq. (\ref{CBC}) using  a quadrature defined in
the half--space, this may introduce  an undesirable mismatch of  the
 nodes of the quadrature used on the boundary and that in  the bulk. Thus, we
here apply the  quadrature used in the bulk even for the boundary
 nodes.   Next, we 
shall estimate the extra error introduced by  this procedure  in
comparison to the discretization error present in the bulk. 

The discrete distribution function used in the LBM is the projection of the
continuous distribution function  in a finite dimensional orthonormal
Hermite basis \cite{ShanHe}. The equilibrium also need to be projected in this
basis to have correct conservation laws. This solution 
  has a major drawback that the positivity of the distribution
  function is lost in the truncation.  
This problem is circumvented, if we evaluate the Boltzmann
$H$--function, rather than its minimizer under the constrains of
conservation of the hydrodynamic variables, for the discrete case. Indeed, 
the  local equilibrium can
 also be written as
\begin{equation}
\label{EQG}
f^{\mbox{eq}}(\mathbf{c},\rho_{\mbox{w}}, U_{\mbox{w}}, T_{\mbox{w}})
= \exp{\left(\alpha+ \mathbf{\beta} \cdot \mathbf{c} + \gamma \mathbf{c}^2\right)},
\end{equation}
where $\alpha$, $\mathbf{\beta}$ and $\gamma$ are the Lagrange multipliers needed
for the minimization of the  Boltzmann $H$--function under the constrains
of conservation of the hydrodynamic variables. These Lagrange
multipliers are calculated from the requirement that the moments of
the  equilibrium distribution $f^{\mbox{eq}}$ are known hydrodynamic
quantities.  
 Now, once we have evaluated  the projection of the  Boltzmann $H$--function on a
 finite dimensional Hermite basis, we calculate the equilibrium from
 the knowledge of the discrete $H$--function. It turns out that the
 equilibrium corresponding to the discrete $H$--function also has  the
 same functional form as Eq. (\ref{EQG}).  Only difference is that the
 Lagrange multipliers has to be calculated from the discrete
 conservation laws. One example of explicit form of such equilibrium distribution
 function  is given in Ref. \cite{AK00}.

 First projecting the distribution functions in the Hermite basis
 and then evaluating
the integrals appearing in  Eq. (\ref{CBC}) by quadrature, we have 
\begin{equation}
\label{DBC}
 \tilde{f}(\mathbf{c}_i, t)
 =  \frac{
 \sum_{\xiprime_i \cdot \mathbf{n} < 0  } 
|(\xiprime_i\cdot\mathbf{n})|
\tilde{f}(\mathbf{c}^\prime_i, t)
}{
\sum_{\xiprime_i \cdot \mathbf{n} < 0  } 
|(\xiprime_i\cdot\mathbf{n})|
\tilde{f}^{\mbox{eq}}(\mathbf{c}^\prime_i, U_{\mbox{w}},  \rho_{\mbox{w}})} 
\tilde{f}^{\mbox{eq}}(\mathbf{c}_i, U_{\mbox{w}},  \rho_{\mbox{w}})
,   \qquad  ( \left( \mathbf{c}_i -\mathbf{U}_{\mbox{w}}\right)\cdot  \mathbf{n} > 0),
\end{equation}
 where 
\begin{equation}
\tilde{f}(\mathbf{c}_i, t) = w_i \rho \frac{f( \mathbf{c}_i, t) }
 {{f}^{\mbox{eq}}(\mathbf{c}_i, 0,\rho)}  
\end{equation}
 denotes the rescaled distribution function 
 evaluated at  nodes of the quadrature. This rescaled distribution
 function is the distribution function used in the LBM \cite{ShanHe,
 Luo1}. The discrete equilibrium distribution function $\tilde{f}^{\mbox{eq}}$ is
 the projection of the equilibrium distribution on the finite
 dimensional Hermite basis and calculated by the procedure discussed
 above.   Before estimating error
 associated with this formula, a few remark about the preceding
 equation is in order.  
First, in the isothermal case the wall temperature is a
redundant quantity and is dropped from the 
argument of equilibrium distribution. 
 To get a boundary condition for the lattice BGK equation, the
 true discrete equilibrium appearing in the Eq. (\ref{DBC}) can be
 replaced by the equilibrium used in the BGK model \cite{Quin}.
This substitution  is justified because up to order $O(M^2)$ the true
 equilibrium can be replaced by the  BGK
 equilibrium.    
However, positivity of  the reflected distributions may be lost in this
 truncation in the same way it happens in the bulk for lattice BGK
 model.  A similar expression for the boundary
 conditions was earlier postulated  by Gatingnol in the context of  
 discrete velocity models of the kinetic theory \cite{Gatin}. 

In order to estimate the extra error introduced on  the boundary 
in comparison  to the
bulk,  we write the ratio of two integrals
appearing in the eq.(\ref{CBC})   as
 \begin{align}
\begin{split}
I 
 &=  \frac{
\int_{\mxi^\prime \cdot \mathbf{n} < 0 } |\mxi^\prime \cdot\mathbf{n}|
f(\mathbf{c}^\prime, t)
  d\mathbf{c}^\prime
}{
\int_{\mxi^\prime \cdot \mathbf{n} < 0 } |\mxi^\prime \cdot\mathbf{n}|
f^{\mbox{eq}}(\mathbf{c}^\prime, \rho_{\mbox{w}}, U_{\mbox{w}}, T_{\mbox{w}}) d \mxi^\prime}  
\\
&= 1+  \frac{
\int_{\mxi^\prime \cdot \mathbf{n} < 0 } |\mxi^\prime \cdot\mathbf{n}|
f^{\mbox{neq}}(\mathbf{c}^\prime, t)
  d\mathbf{c}^\prime
}{
\int_{\mxi^\prime \cdot \mathbf{n} < 0 } |\mxi^\prime \cdot\mathbf{n}|
f^{\mbox{eq}}(\mathbf{c}^\prime, \rho_{\mbox{w}}, U_{\mbox{w}}, T_{\mbox{w}}) d
 \mxi^\prime}, 
\end{split}
\end{align}
 where $\ff^{\mbox{neq}}= \ff- \ff^{\mbox{eq}}$. As discussed above,  the evaluation of the integral appearing  in the
 denominator  is straightforward in the sense
 that the order of accuracy of this evaluation is same as that of
 moments evaluation in the bulk.  
 In order to evaluate the integral appearing in the numerator,
  we 
 perform Hermite expansion of the non--equilibrium part of the
 distribution function around the zero velocity equilibrium. The
 result is

\begin{align}
f^{\mbox{neq}}(\mathbf{c},t) &= f^{\mbox{eq}}(\mathbf{c}, \rho_{\mbox{w}}, 0, T_{\mbox{w}} ) 
\sum_{i=0}^{N} \frac{a^{(i)}}{i!} {\cal{H}}^{(i)} \left( \mathbf{c} \right).
\end{align}
The first two expansion
coefficients of the non--equilibrium part of the population  are   $(a^{(0)}=0, a^{(1)}_i=0$).
In case of the isothermal hydrodynamics, only non--zero Hermite
coefficient needs to be kept is 
$a^{(2)}_{i j}$. This is a symmetric  tensor and is
independent of the particle velocity $\mathbf{c}$. After using
the symmetry of the second order Hermite--polynomials,
 \begin{align}
I 
&= 1+ \frac{1}{2}\frac{a_{\alpha \beta }^{(2)}
\int |\mxi^\prime_i \cdot\mathbf{n}|f^{\mbox{eq}}(\mathbf{c}^\prime, \rho_{\mbox{w}}, 0, T_{\mbox{w}} ) 
{\cal{H}}_{\alpha \beta}^{(2)}(\mathbf{c}^\prime)
  d\mathbf{c}^\prime
}{
\int_{\mxi^\prime \cdot \mathbf{n} < 0
}  |\mxi^\prime \cdot\mathbf{n}|
f^{\mbox{eq}}(\mathbf{c}^\prime, \rho_{\mbox{w}}, U_{\mbox{w}}, T_{\mbox{w}}) d
 \mathbf{c}^\prime}. 
\end{align}
 This expression can be evaluated using the Gauss--Hermite quadrature. The result is
\begin{align}
I 
&= 1+ \frac{
\sum_{
\mxi^\prime_i \cdot \mathbf{n} < 0
 }
 a_{\alpha \beta}^{(2)}
w_i |\mxi^\prime_i \cdot\mathbf{n}|
{\cal{H}}_{\alpha \beta }^{(2)}(\mathbf{c}^\prime_i)
}{
\sum_{\mxi^\prime_i \cdot \mathbf{n} < 0
} |\mxi^\prime_i \cdot\mathbf{n}|
\tilde{f}^{\mbox{eq}}(\mathbf{c}^\prime, \rho_{\mbox{w}}, U_{\mbox{w}})}. 
\end{align}
This expression gives an estimate of the order of the accuracy of the
  eq.(\ref{DBC}).  In  evaluation of the moments ( up to the
  second order moment)  of the distribution
  function, no extra error is introduced as compared to the
  bulk. This happened  because first odd order Hermite coefficient
  appearing in the expansion is zero.   
  Due to the expansion
around global equilibrium, used in the derivation, the
boundary condition is valid only up to the order $O(M^2)$.

Now, for purely diffusive scattering, we have a closed form expression
for the reflected populations with the same order of accuracy as the
bulk node. However, we have said nothing about the
grazing directions.  Unlike continuous kinetic theory, here we need to
specify the conditions in the grazing directions. Only information we
have about the grazing populations is their positivity. A simple
way to fixed the grazing population is to let them evolve
according to the lattice--Boltzmann equation like nodes in the bulk
region. This condition is implemented in the simulations presented in
the next section. 

\section{ Numerical tests}
\label{Sec:4}
The boundary condition derived in the previous section (Eq. (\ref{DBC})), retains one important
feature of original Boltzmann equation, the Knudsen number dependent
slip at the wall. To show this, we have 
performed a numerical simulation of the Kramers' problem 
\cite{CBoltz}. This is one of the few problem where solution of the
continuous Boltzmann equation  is known
analytically. This problem is a limiting case of the plane Couette flow,
where one of the plate is moved to infinity, while keeping a fixed
shear rate. We compare the analytical solution for the slip--velocity
at the wall calculated for the  linearized BGK collision
model with the numerical solution in the Fig. \ref{Fig1}. We have
performed the numerical computation for the D2Q9 lattice with the
entropic formulation of the LBM \cite{AK00, AK01} with the expression
of the $H$ function given by Eq. (\ref{app:H}). The
agreement between the two result for Knudsen number going to zero is
very good. This is indeed an important result as it shows that
with the proper implementation of the boundary condition, the solution
of the LBM converge to the hydrodynamic limit (Knudsen number going to
zero) in the same way as the Boltzmann equation.

 By simulating the Kramer's problem, we have shown that the present
 boundary condition can be used for stationary wall. To validate the
 boundary condition for the moving wall, we have performed simulation
 of the lid--driven cavity flow. The plot of stream--function is given
 in Fig.\ref{Fig2} for Reynolds number $Re=1000$. The location of the
 primary and secondary vortex
 and the magnitude of the stream--functions agrees well with the previous 
simulations \cite{Hou}.

  Once we have shown that the diffusive boundary condition used for
 the continuous Boltzmann equation can be reformulated for the discrete
case, the question arises that, can this procedure be applied  for a more
 sophisticated scattering kernels used in the continuous kinetic
 theory (see for  examples Ref. \cite{CBoltz}). The answer is in affirmative for any condition written
 in the integral form, while in general it cannot be done for a point-wise
 condition like purely specular reflection. For example, a  very general form of
 scattering probability, written in the integral form and 
 can be easily modified  for nonzero mass flux,  is given in the
 Ref. \cite{CBoltz} (see Eq.($6.26$) of the Chapter $3$). This form
 of the scattering probability can be
 discretized using the present method.

It is instructive to compare 
the `bounce--back' condition used in the literature with the present
boundary condition. It can be seen easily that the present boundary
condition reduces to the bounce--back condition for the three velocity
model used in the one--dimensional
case. However, there is no correspondence between the two condition in
the higher dimensions. 

 The present boundary condition retains
the positivity at the boundary nodes. This is a major advantage in
comparison  to other proposed boundary conditions for the purpose of the
numerical stability.
The Knudsen number dependent wall slip is a manifestation of the
kinetic nature of the lattice Boltzmann equation.  This nature of the
scheme can be a burden if one is interested in solving the  macroscopic
creeping flow problems with very small grid size. This will put some
restriction on  the
simulation of creeping flow in  very small grids (lattice Knudsen
number $= \nu / L c_s$).  However, the restriction is not  severe
because of the fact that we still have the freedom to choose velocity
very small to attain zero Reynolds number situation. In fact, the
same condition is required for the validity of the LBM simulation of
the hydrodynamics in the bulk.  
 To conclude,  we have proposed 
boundary conditions based on the kinetic theory considerations. A
systematic way of dealing with the conditions at the boundary is
developed for the lattice--Boltzmann method.  The present work opens the
way to the future development for the cases of reactive, porous or
adsorbing walls. 

\section{Acknowledgements}
  We thank Professor Hans Christian \"Ottinger for the useful discussions.
  
  \appendix
\begin{appendix}
\section{\label{app:1}Derivation of the $H$--function}
The Boltzmann $H$--function is 
\begin{equation}
H = \int f \log f d\mxi
\end{equation}
This expression written in terms of the logarithm of the
distribution function $(\mu = \log{f})$ is 
\begin{equation}
H = \int \mu \exp{(\mu)} \; d\mxi
\end{equation}
 We have chosen to work with the variable $\mu$  because the 
 projection  of it on to  the  Hermite basis preserves  positivity of the
 distribution function $f$. The expansion of the function $\mu$ is
\begin{equation}
\mu= 
A^{(0)} {\cal{H}}^{(0)} \left( \mathbf{c} \right)+
A^{(1)}_\alpha {\cal{H}}^{(i)} \left( \mathbf{c} \right) +
A^{(2)}_{\alpha \beta} {\cal{H}}^{(i)}_{\alpha \beta} \left( \mathbf{c} \right)
\end{equation}
This expression can also be written as,
\begin{equation}
\mu= 
A^{(0)} {\cal{H}}^{(0)} \left( \mathbf{c} \right)+
A^{(1)}_\alpha {\cal{H}}^{(i)} \left( \mathbf{c} \right) +
B^{(2)}_{\alpha \beta} {\cal{H}}^{(i)}_{\alpha \beta} \left(
  \mathbf{c} \right)
- \frac{\mathbf{c}^2}{2}
\end{equation}

The expansion coefficients $A$ is calculated by the requirement that
the moments of $\exp{(\mu)}$ are hydrodynamic variables. The expansion
used here is  a  slightly different form of the Grad's moment expansion \cite{Grad2}
and is known as the maximum entropy approximation \cite{Kogan65, ANG,Levermore96}.
Now, the Boltzmann $H$ function is  in a integral form suited for the
evaluation in the Gauss--Hermite quadrature (see
Eq. (\ref{GHQ})). This evaluation gives the discrete form of the $H$
function as
\begin{equation}
\label{app:H}
H=\sum_{i=1}^{b}\tilde{f}_{i}\ln\left(
\frac{\tilde{f}_{i}}{w_i} \right),
\end{equation}
 where,
\begin{equation}
\tilde{f}(\mathbf{c}_i, t) = w_i (2 \pi k_B T_0)^{D/2} \exp{\left(\frac{\mathbf{c}^2_i}{2}
  \right) } f( \mathbf{c}_i, t), 
\end{equation}
where $T_0$ is the reference temperature. 
 In $2$--dimension, the nodes of the quadrature and the corresponding weights are
\begin{align}
\mathbf{c}_i =\begin{cases}
 \{0,0 \}  & \text{ if $i =0$}\\
 c \left \{
(\cos\left(  \frac{\pi(i-1)}{2}\right) , \sin
 \left(\frac{\pi(i-1)}{2}\right)\right \}       & \text{ if $i =1,2,3,4$}\\
 c \sqrt{2} \left \{
(\cos\left(  \frac{\pi(2  i-9)}{4}\right) , \sin
 \left(\frac{\pi(2  i - 9)}{4}\right)
 \right \}   &  \text{ if $i =5,6,7,8$},\\
\end{cases}
\end{align} 
and

\begin{align}
w_i =\begin{cases}
 \frac{4}{9}  & \text{ if $i =0$}\\
 \frac{1}{9}       & \text{ if $i =1,2,3,4$}\\
\frac{1}{36}      &  \text{ if $i =5,6,7,8$}.\\
\end{cases}
\end{align}

Here the magnitude of the discrete velocity $c$ is related to the
reference temperature by the relation $c = \sqrt{(3 k_B T_0)}$.
With this, the entropy expression derived here coincide with the
expression derived in  Ref. \cite{K99}, by a
different argument.  
\end{appendix}

\begin{figure}[ht]
 \begin{center}
\caption{\label{Fig1}  Relative slip observed at the wall in the simulation of the
  Kramers' problem for 
shear rate   $a = 0.001 $,  box length $L=32$, $v_\infty = a \times L
= 0.032$. All the quantities are given in the dimensionless lattice unit. 
}
 \end{center}
\end{figure}

\begin{figure}
 \begin{center}
\caption{\label{Fig2} Stream--function for $Re=1000$ in a simulation of lid driven
  cavity flow.  Parameters used are:  grid size $320 \times 320$, and 
 lid velocity   $V = 0.075 $.   All the quantities are given in the
 dimensionless lattice unit.}
 \end{center}
\end{figure}

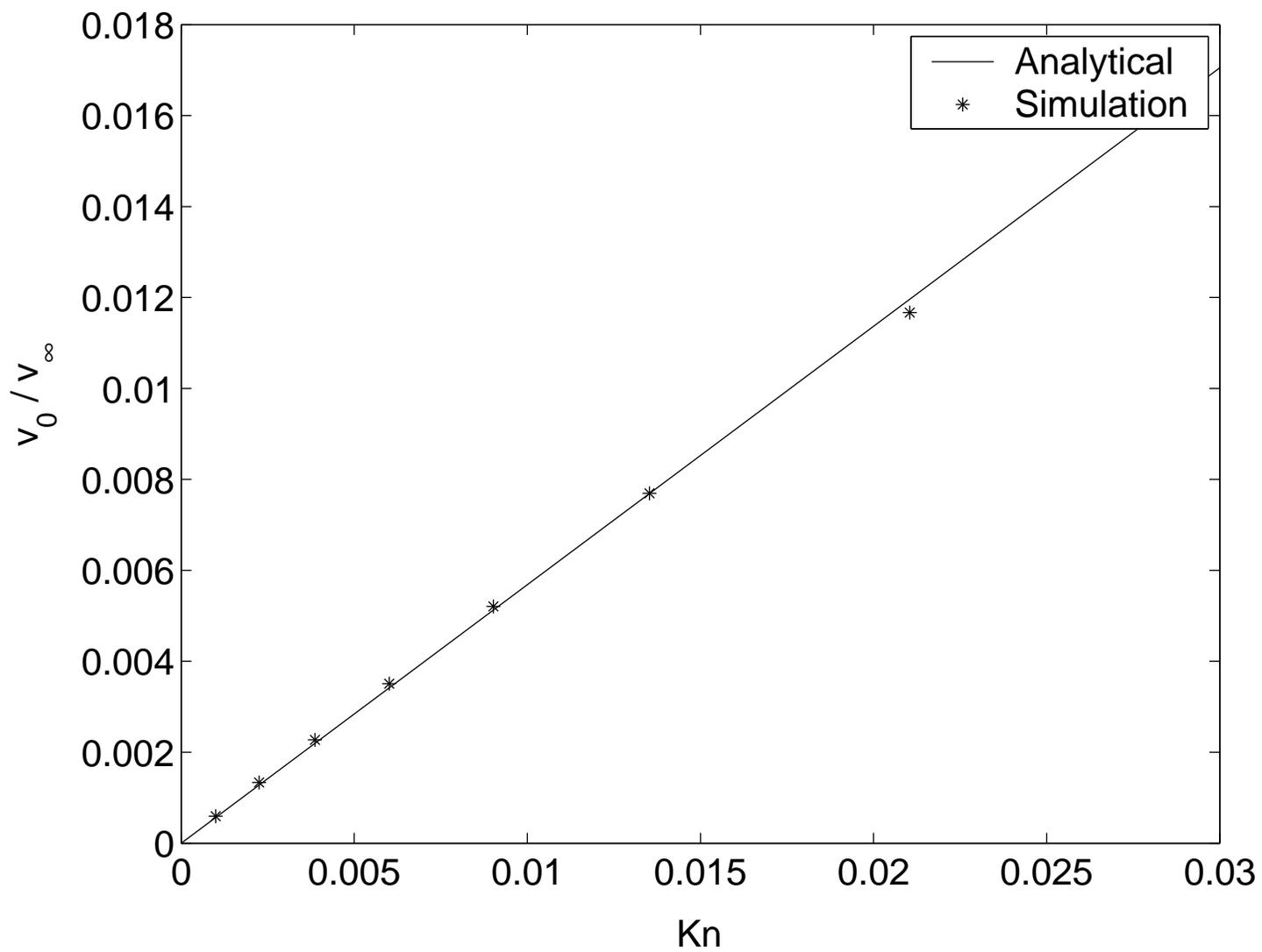
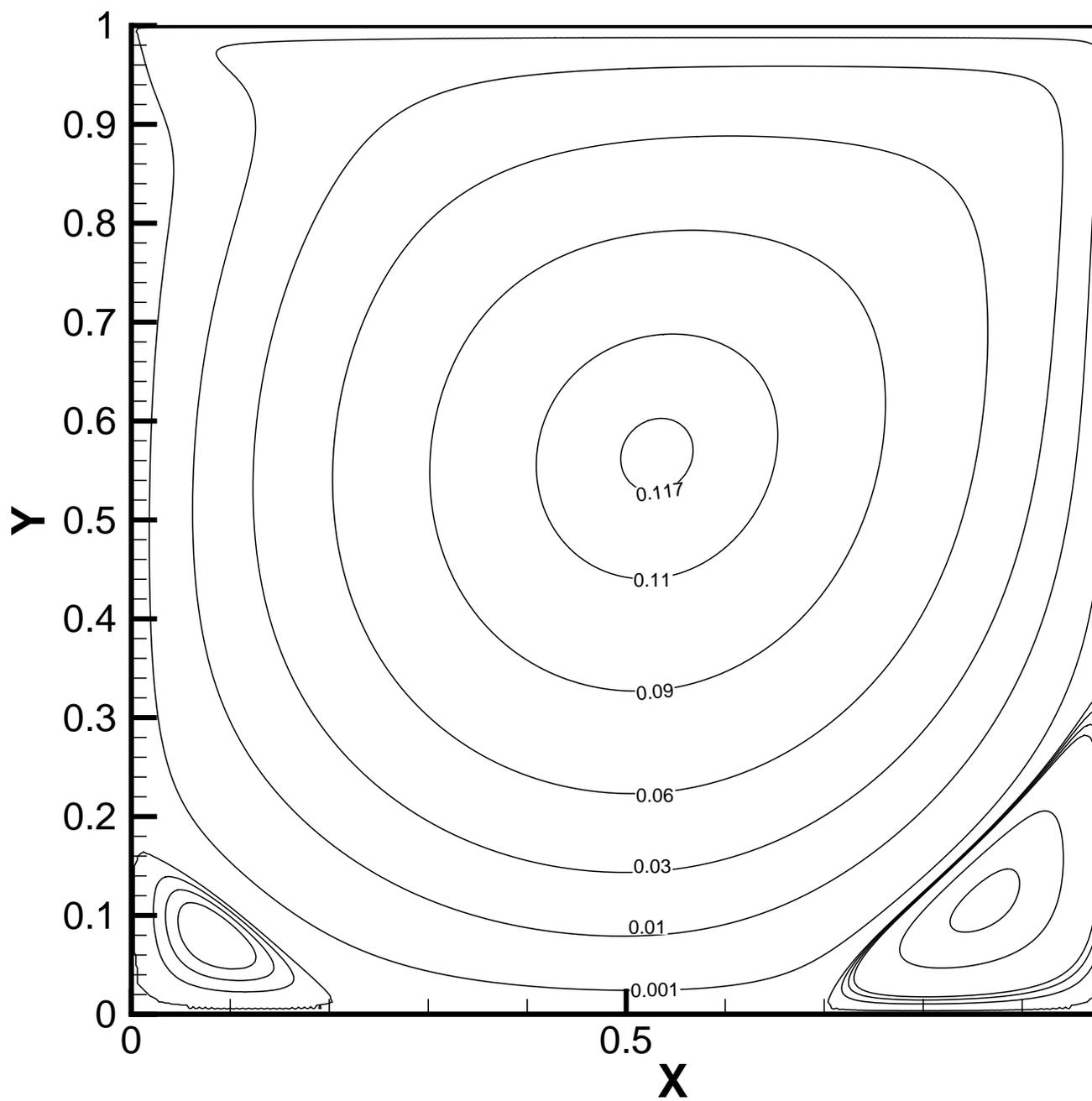
\end{document}